
\input phyzzx
\def\D{\Delta}
\def\cm{{\cal{M}}}
\def\ck{{\cal{K}}}
\def\gt{{\tilde g}}
\def\Bt{{\tilde B}}
\def\Pht{{\tilde\Phi}}

\NPrefs
\def\define#1#2\par{\def#1{\Ref#1{#2}\edef#1{\noexpand\refmark{#1}}}}
\def\con#1#2\noc{\let\?=\Ref\let\<=\refmark\let\Ref=\REFS
         \let\refmark=\undefined#1\let\Ref=\REFSCON#2
         \let\Ref=\?\let\refmark=\<\refsend}


\define\GIBB
G. Gibbons and K. Maeda, Nucl. Phys. {\bf B298} (1988) 741;
D. Garfinkle, G. Horowitz and A. Strominger, Phys. Rev. {\bf D43}
(1991) 3140;
G. Horowitz and A. Strominger, Nucl. Phys. {\bf B360} (1991) 197;
A. Shapere, S. Trivedi and F. Wilczek, Mod. Phys. Lett.
{\bf A6} (1991) 2677.

\define\GIDD
S. Giddings and A. Strominger, Phys. Rev. Lett.
{\bf 67} (1991) 2930.

\define\MYERS
R. C. Myers, Nucl. Phys. {\bf B289} (1987) 701;
R. C. Myers and M. Perry, Ann. Phys. {\bf 172} (1986) 304;
C. Callan, R. C. Myers and M. Perry, Nucl. Phys. {\bf B311}
(1988) 673.

\define\DABH
A. Dabholkar, G. Gibbons, J. Harvey anf F. R. Ruiz,
Nucl. Phys. {\bf B340} (1990) 33.

\define\CALL
C. Callan, J. Harvey and A. Strominger, Nucl. Phys. {\bf B359}
(1991) 611.

\define\DUFF
M. Duff and J. Lu, Phys. Rev. Lett. {\bf 66} (1991) 1402;
Nucl. Phys. {\bf B354} (1991) 141.

\define\MSW
G. Mandal, A. M. Sengupta and S. Wadia, Mod. Phys. Lett. {\bf A6}
(1991) 1685.

\define\WITT
E. Witten Phys. Rev. {\bf D44} (1991) 314;
C. Nappi and E. Witten, IASSNS-HEP-92-38;
J. Horne and G. Horowitz, Nucl. Phys. {\bf B368} (1992) 444.

\define\SEN
A. Sen, TIFR/TH/92-57 (hep-th/9210050);
G. Horowitz, UCSBTH-92-32 (hep-th/9210119);
J. Harvey and A. Strominger, EFI-92-41 (hep-th/9209055).

\define\ASEN
A. Sen, Phys. Rev. Lett. {\bf 69} (1992), 1006.



\define\VENE
G. Veneziano, Phys. Lett. {\bf B265} (1991) 287;
K. Meissner and G. Veneziano, Phys. Lett. {\bf B267} (1991) 33;
M. Gasperini, J. Maharana and G. Veneziano, Phys. Lett. {\bf B272}
(1991) 277, CERN-TH-6634/92 (hep-th/9209052).

\define\ASHOKE
A. Sen, Phys. Lett. {\bf B271} (1991) 295;
{\it{ibid}} {\bf B274} (1991) 34.

\define\HASSAN
S. F. Hassan and A. Sen, Nucl. Phys. {\bf B375} (1992) 103.

\define\MICRO
A. Sen, TIFR/TH/92-29, hep-th/9206016.

\define\KKK
S. Kar, S. Khastgir and A. Kumar, Mod. Phys. Lett. {\bf A7}
(1992) 1545;
S. Khastgir and A. Kumar, Mod. Phys. Lett. {\bf A6} (1991) 3365;
S. Kar and A. Kumar, IP-BBSR-92-18;
J. Maharana, CALT-68-1781;
S. Khastgir and J. Maharana, IP-BBSR-92-38 (hep-th/9206017);
A. Kumar, Phys. Lett. {\bf B293} (1992) 49;
P. Horava, Phys. Lett. {\bf B278} (1992) 101;
J. Panvel, Phys. Lett. {\bf B284} (1992) 50;
S. F. Hassan and A. Sen, TIFR/TH/92-61 (hep-th/9210121);
S. Khastgir and J. Maharana, IP-BBSR-92-77 (Nov. 1992).

\define\GIVEON
A. Giveon and A. Pasquinucci, Phys. Lett. {\bf B294} (1992) 162;
A. Giveon and M. Rocek, Nucl. Phys. {\bf B380} (1992) 128.

\define\ROCEK
M. Rocek and E. Verlinde, Nucl. Phys. {\bf B373} (1992) 630.


\define\MAHARANA
J. Maharana and J. Schwarz, CALT-68-1790.

\define\HORNE
J. Horne and Horowitz, Phys. Rev. {\bf D46} (1992) 1340.

\define\HHS
J. Horne, G. Horowitz and A. Steif, Phys. Rev. Lett. {\bf 68}
(1992) 568.

\define\BUSH
T. Busher, Phys. Lett. {\bf B201} (1988) 466;
{\it{ibid}} {\bf B194} (1987) 59.

\define\FROLOV
V. P. Frolov, A. I. Zelnikov and U. Bleyer, Ann. Phys. {\bf 44}
(Leipzig),
(1987) 371;
G. Gibbons and D. Wiltshire, Ann. Phys. {\bf 167} (1986) 201; {\bf 176}
(1987) 393(E).

\define\HORO
G. Horowitz and A. strominger, Nucl. Phys. {\bf B360} (1991) 197.

{}~\hfill\vbox{\hbox{IMSC/93- 6}\hbox{January, 1993}}\break

\title{ON THE ROTATING CHARGED BLACK STRING SOLUTION}

\author{Swapna Mahapatra
\foot{ e-mail: swapna@imsc.ernet.in}}
\address{ Institute of Mathematical Sciences,\break
C. I. T. Campus, Madras-600113, India}

\abstract
A rotating charged black string solution in the low energy
effective field theory describing five dimensional heterotic
string theory is constructed. This solution is labelled by mass,
electric charge, axion charge and angular momentum per unit length.
The extremal limit of this solution is also studied.

\endpage

The construction of classical solutions in
string theory has received much attention in recent
times \con\GIBB\GIDD\MYERS\DABH\CALL\DUFF\MSW\noc.
The study of these solutions has shown the
existence of black hole and extended black holes $\it{i. e.}$
black strings and black p-brane type structures. Solvable
conformal field theories are known in some cases \WITT.
These black holes and extended objects in string theory are very
different from those which occur in general relativity because
of the presence of the nontrivial dilaton field. In Einstein-
Maxwell theory, all stationary black holes are described
by the Kerr-Newman solutions parametrized by three quantities,
namely the mass $M$, charge $Q$ and the angular momentum
parameter $a$. When $a = 0$, the solution describes the charged
black hole solution or the Reissner-Nordstrom solution in
general relativity. When $Q = 0 = a$, the solution reduces
to that of the Schwarzschild solution. The vacuum Kerr family
of solutions have $Q = 0$ and they describe the rotating
uncharged black hole solution. The generalizations of all these
solutions to higher dimensions have also been discussed in the
literature.

The study of these classical solutions in string theory is
important as string theory is expected to provide us with a
consistent quantum theory of gravity and also it will help us
in understanding the basic nature of string theory itself. There
exists a vast literature in this area of research and we shall not
go into the details of it. For recent review, see ref. \SEN.
In this paper, we shall only concentrate on the rotating charged
black string solution which can appear in the low energy heterotic
string theory. Rotating charged black hole solutions in string theory
have been obtained by Sen \ASEN. These solutions have been obtained
from the Kerr solution in general relativity by using the method
of twisting \con\VENE\ASHOKE\HASSAN\KKK\MAHARANA\MICRO\GIVEON
\ROCEK\noc. The method of twisting basically means that in
string theory, if we have an exact classical solution which is
independent of $d$ of the space-time coordinates, then we can
perform an $0(d) \times 0(d)$ transformation on the solution, which
produces new inequivalent classical solutions satisfying the same equation
of motion derived from the low energy string effective action.
Similarly in the case of heterotic string theory, the space
of classical solutions which are independent of $d$ of the space-
time directions and for which the gauge field configuration
lies in a subgroup that commutes with $p$ of the $U(1)$ generators
of the gauge group, has an $0(d) \times 0(d + p)$ or more generally
$0(d - 1, 1) \times 0(d + p -1, 1)$ symmetry and  using this
transformation, one can generate new inequivalent classical solutions
starting from the known ones \HASSAN. Various interesting solutions
have been constructed by using this twisting method and the
application of this solution has been widely studied [10-18].
The same philosophy has been used in ref. \ASEN\ to generate
the rotating charged black hole solution in heterotic string theory
starting from the Kerr solution.

It is also important to study the black string solution in order
to have a deeper understanding of the string theory itself. Black
strings are one dimensional extended objects surrounded by event
horizons. There exists a rich variety of extended black hole solutions
in ten dimensional string theory and they are closely related to the
string soliton solution as well as to the fundamental strings
\HORO.
These solutions are labelled by mass and axion charge per unit length.
Rotating black string solutions have also been obtained by Horne and
Horowitz \HORNE, where the solution is parametrized by mass, axion charge,
and angular momentum per unit length. However, they have not included
the Maxwell field in the low energy effective action. In this paper,
we construct the most general rotating charged black string solution
carrying mass,
electric charge, axion charge and angular momentum in the five dimensional
low energy field theory describing heterotic string theory.
Once again we use the method of twisting to obtain this solution starting
from a four dimensional Kerr solution with an extra flat direction.
We study the extremal limit of this solution to know
the behaviour of extremal black string when both charge and rotation
are present.
We find that angular momentum dominates over the charge in the extremal
limit which is also a characteristic feature of rotating charged black hole
solution.

We start with the low energy action for heterotic string theory in five
dimensions. Apart from the five dimensional string metric, we have dilaton,
antisymmetric tensor gauge field and the Maxwell field. We do not consider
the massless fields arising due to compactification and also we retain
only terms with two or less number of derivatives in the action.
Such an action is given by,

$$S = \int d^5x {\sqrt{-g}} e^{-2\Phi} ( -R - 4 {(\nabla\Phi)}^2 +
{1\over 12}H^2 + {1\over 2} F^2) \eqn\one$$

Here, $g_{\mu\nu}$ and $\Phi$ denote the metric and the dilaton field.
$R$ is the five dimensional Ricci scalar and $F_{\mu\nu} =
\partial_{\mu} A_{\nu} - \partial_{\nu} A_{\mu}$, is the field strength
corresponding to the $U(1)$ gauge field $A_{\mu}$. The three form $H$
is given by,

$$H_{\mu\nu\rho} = \partial_{\mu} B_{\mu\nu} + {\rm {cyclic~ permutations}}
- (\Omega_3(A))_{\mu\nu\rho}, \eqn\two$$

 where, $B_{\mu\nu}$ is the antisymmetric tensor gauge field and
 $(\Omega_3(A))_{\mu\nu\rho}$ is the gauge Chern-Simons term. The
 Lorentz C-S term has been neglected as they involve more than two
 derivatives in the action.

 Given a solution $G_{\mu\nu}$, $B_{\mu\nu}$, $\Phi$ and $A_{\mu}$,
 of the classical equation of motion, we would like to obtain the new
 transformed solution which also
 satisfies the same equations of motion derived from the action (1), by
 using the twisting procedure. For this purpose, we define an $11 \times 11$
 matrix $\cm$ as \HASSAN,

$$
{\cm} =\pmatrix{({\ck}^T - \eta) g^{-1} ({\ck} - \eta)&
({\ck}^T - \eta) g^{-1} ({\ck} + \eta)& - ({\ck}^T - \eta)
g^{-1} A\cr
({\ck}^T + \eta) g^{-1} ({\cal K} - \eta)&
({\cal K}^T + \eta) g^{-1} ({\cal K} + \eta)
& -({\cal K}^T + \eta) g^{-1} A \cr
- A^T g^{-1}({\cal K} - \eta)& -A^T g^{-1} ({\cal K} + \eta)
& A^T g^{-1} A
\cr}\eqn\three$$

where,

$${\cal K}_{\mu\nu} = - B_{\mu\nu} - g_{\mu\nu} -
{1\over 4} A_{\mu} A_{\nu}\eqn\four$$

and,

$$\eta_{\mu\nu} = {\it{diag}}\,(1, 1, 1, 1, -1)\eqn\five$$

Here, $g_{\mu\nu}$ and $B_{\mu\nu}$ are $5 \times 5$ matrices and $A_{\mu}$
is a five dimensional column vector. So given the solution of the equations
of motion (as derived from action (1)), one can generate the new inequivalent
solution, where the two solutions are related through the following
relations:

$${\cm}'= \Omega \,\cm \,{\Omega}^T;\qquad
{\Phi}' - {1\over 2} ln\sqrt{det\, g'} = \Phi - {1\over 2}
ln\sqrt{det\, g},\eqn\six$$

where, ${\cm}'$ is the same matrix as $\cm$ but with the new
variables $g'_{\mu\nu}$, $B'_{\mu\nu}$ and $A'_{\mu}$. $\Omega$
is an $O(5, 6)$ matrix satisfying,

$$\Omega \,L \Omega^T = L\eqn\seven$$

where,
$$L = {\it{diag}}\,(\eta_5, -\eta_6)\eqn\eight$$.

We want to generate the electrically charged rotating black string
solution as an application of this transformation.
Basically one starts
with the four dimensional Kerr solution describing the rotating
black hole and adds one extra dimension to it.
The corresponding metric is given by,

$$\eqalign{ds^2 &= - {(\D - a^2 \sin^2\theta)\over{\Sigma}} dt^2 -
{2 a \sin^2\theta(r^2 + a^2 - \D)\over{\Sigma}}dt d\phi \cr
&+
[{(r^2 + a^2)^2 - \D a^2 \sin^2\theta\over{\Sigma}}]
\sin^2\theta d\phi^2 + {\Sigma\over\Delta} dr^2 + \Sigma d\theta^2 + dx^2,
\cr}\eqn\nine$$

where,
$\D = r^2 + a^2 - 2 m r$; $\Sigma = r^2 + a^2 \cos^2\theta$;
and $a$ is the angular momentum parameter.

We also have,

$$\Phi = 0; \quad B_{\mu\nu} = 0; \quad A_{\mu} = 0.\eqn\twelve$$

In order to obtain the black string solution with non zero axion charge,
one Lorentz boosts the solution to produce a non zero linear momentum
along the $x$ (extra flat) direction and then uses the sigma model
duality relations to convert this momentum to charge \HHS. This is a
novel way of adding axion charge to a static and translationally
invariant solution.
One knows that given a solution $g_{\mu\nu}$, $B_{\mu\nu}$
and $\Phi$ with a translational symmetry in $x$ direction, the dual
solution is obtained from the following relations \BUSH,

$$\eqalign{
\gt_{xx}= {1\over g_{xx}}; &\qquad \gt_{x \alpha}=
{B_{x \alpha}\over g_{xx}};\cr
\gt_{\alpha\beta}= g_{\alpha\beta} &- {(g_{x\alpha} g_{x\beta} -
B_{x\alpha} B_{x\beta})\over{g_{xx}}};\cr
\Bt_{x\alpha}= {g_{x\alpha}\over{g_{xx}}};& \qquad \Bt_{\alpha\beta}=
B_{\alpha\beta} - {2 g_{x[\alpha} B_{\beta]x}\over{g_{xx}}};\cr
{\Pht} = \Phi\, + \,&{1\over2}\, {\rm log}~ g_{xx}}\eqn\thirtn
$$

In fact this was the approach taken by Horne and Horowitz to construct
the rotating black string solution with a nonzero axion charge. Their
solution is given by,

$$
\eqalign{d s^2 &= - {(1 - Z)\over B^2} d t^2 -
{2 a Z \sin^2\theta\over
{B^2 \sqrt{1 - v^2}}} d t\, d{\phi} \cr &+
[(r^2 + a^2) + a^2 \sin^2\theta {Z\over{B^2}}]\sin^2\theta\,d{\phi}^2
+ {\Sigma\over \D}dr^2 \cr &+
\Sigma d{\theta}^2 + B^{-2} d x^2,\cr}\eqn\fortn
$$

where, $Z = {2 m r\over \Sigma}$.
The dilaton and the non zero components of the antisymmetric tensor
field are given by,

$$\Pht = - {\rm log}~ B,\eqn\fiftn$$

$$\Bt_{x t} = {v\over{1 - v^2}} {Z\over B^2}; \qquad \Bt_{x\phi} =
- {a Z v \sin^2\theta \over{B^2\,\sqrt{1 - v^2}}}\eqn\sixtn$$

Here $v$ is the boost velocity.
This solution is very much similar to the Kaluza-Klein black hole,
which was obtained for the dilaton coupling parameter
$\alpha = \sqrt 3$ \FROLOV. We can always perform an $O(d, d)$
transformation on this electrically neutral solution to obtain the
charged rotating black string solution. But instead of doing this,
we shall perform the $O(d, d)$ transformation on the four dimensional
Kerr solution with an extra flat direction (eqn. 9), which will
automatically generate the rotating black string solution with a nonzero
electric as well as axionic charge.
In order to obtain the
inequivalent field configurations, we consider a mixing between
the coordinates $t$, $x$ and one of the internal coordinate
$y$ corresponding to one of the non abelian gauge field included
in the action. So the most general transformed solution is obtained by first
considering a boost in the $t - y$ ($y$ is the coordinate in
the internal space) direction, followed by a boost in the $t -
x$ direction. The matrix $\Omega$ is given by,

$$\Omega = \pmatrix{I_3&&&\cr &S&&\cr &&I_3&\cr
&&&R\cr}\eqn\seventn$$

where $S$ and $R$ are $O(1, 1)$ and $O(2, 1)$ matrices associated
with the Lorentz transformation respectively. We choose the matrix
$S$ to be identity matrix. The matrix $R$ is chosen to be,

$$R = \pmatrix{\cosh\alpha_2&\sinh\alpha_2&0 \cr
\sinh\alpha_2&\cosh\alpha_2&0 \cr
0&0&1 \cr} \pmatrix{\cosh\alpha_1&0&\sinh\alpha_1 \cr
0&1&0 \cr
\sinh\alpha_1&0&\cosh\alpha_1 \cr},\eqn\eitn$$
where, $\alpha_1$ and $\alpha_2$ are arbitrary parameters.
So the matrix $\Omega$ is given by,

$$\Omega = \pmatrix{I_8&\cr
&\cosh\alpha_1\cosh\alpha_2&\sinh\alpha_2&\cosh\alpha_2\sinh\alpha_1 \cr
&\sinh\alpha_2\cosh\alpha_1&\cosh\alpha_2&\sinh\alpha_1\sinh\alpha_2 \cr
&\sinh\alpha_1&0&\cosh\alpha_1\cr}\eqn\nintn$$

With this choice of $\Omega$, we determine the transformed solutions
from the relation,
${\cm}' = \Omega \cm {\Omega}^T$.
The transformed metric is given by,

$$
\eqalign{d {s'}^2 &= - {[\Sigma(\D - a^2 \sin^2\theta) - \beta^2
m^2 r^2]\over{{[\Sigma - m r (1 - A)]}^2}}d t^2 + {2 \beta m r\over{\Sigma
- m r (1 - A)}}d x d t +
{2 \beta m r a \sin^2\theta\over{\Sigma - m r (1 - A)}}d x d \phi \cr
&-{2 m r a \sin^2\theta[(1 + A)\Sigma + \beta^2 m r]\over{{[\Sigma
- m r (1 - A)]}^2}}d t d \phi
+ \{[{(r^2 + a^2)^2 - \D a^2 \sin^2\theta\over\Sigma}]\sin^2\theta
\cr
&+ {m^2 r^2 a^2
\sin^4\theta\over\Sigma}[{2(1 - A)\over\Sigma - m r(1 - A)} - {\gamma^2
\Sigma\over{{(\Sigma - m r (1 - A))}^2}}]\}d \phi^2 \cr
&+ {\Sigma\over\D}d r^2 + \Sigma d \theta^2 + d x^2 \cr}\eqn\twentione$$

The nonzero components of the gauge field and the antisymmetric
tensor field are given by,

$$\eqalign{A'_t&= {2 m r \gamma\over{\Sigma - m r(1 - A)}};\cr
A'_{\phi}&= {- 2 m r \gamma a \sin^2\theta\over{\Sigma - m r (1 - A)}};\cr
A'_x&= 0
\cr}\eqn\twentyone$$

and,

$$\eqalign{B'_{t x} &= {\beta m r\over{\Sigma - m r (1 - A)}};\cr
B'_{\phi x} &= {- m r a \sin^2\theta\over{\Sigma - m r (1 - A)}};\cr
B'_{\phi t} &= {m r a (1 - A)\sin^2\theta\over{\Sigma - m r (1 - A)}}
\cr}\eqn\twentytwo$$.
Here,

$$\beta = \sinh\alpha_2 \cosh\alpha_1; \quad \gamma = \sinh\alpha_1;
\quad A = \cosh\alpha_1 \cosh\alpha_2 ,\eqn\twenttwo$$
satisfying the relation $A^2 = 1 + {\beta}^2 + {\gamma}^2$.
The dilaton field is given by,

$$\Phi' = - log\,{[1 - {m r\over\Sigma}(1 - A)]}^{1/2}
\eqn\twentythr$$

{}From the above expressions, we see that in the limit $\alpha_1$,
$\alpha_2  \rightarrow 0$, the solution reduces to that of the
Kerr solution with a flat direction. In the limit
$\alpha_2 \rightarrow 0$, this solution reduces to that of the
rotating charged black hole solution of Sen, with an extra flat direction
and with a replacement of $\Phi \rightarrow 2\Phi$.
In the limit when $a$ $\rightarrow 0$, this solution exactly matches
with the black string solution
obtained by Hassan and Sen when we compactify
five of the flat coordinates \HASSAN. The Einstein
metric is obtained by multiplying expression (20)
with $e^{-{4\over 3} \Phi'}$, {\it{i.e.}}
${g'^E_{\mu\nu}} = e^{-{4\over 3} \Phi'} {g'^{\Sigma}_{\mu\nu}}$.
We shall not here give the complicated expression for the Einstein
metric. This solution has both an event horizon and an inner horizon
at $r_{\pm} = m \pm \sqrt{m^2 - a^2}$ respectively.
The physical mass per unit length is computed by using the ADM mass
formula and the expression is given by,

$$M = {m\over 2}(1 + A)
\eqn\twentyfour$$

The electric charge $Q$ and the magnetic moment $\mu$ are
determined from the asymptotic form of $A_t$ and $A_{\phi}$
respectively and the corresponding expressions are given by,

$$\eqalign{Q &= 2 m \sinh\alpha_1;\cr
\mu &= 2 m a \sinh\alpha_1.\cr}\eqn\twentyfive$$

The expression for the angular momentum $J$ is obtained by
knowing the asymptotic form of the component $g'_{t\phi}$
of the metric and is given by,

$$J = {1\over 2}m a (1 + A)\eqn\twentysix$$

The gyromagnetic ratio is obtained from the standard expression,

$$g = {2 \mu M\over{Q J}} = 2 \eqn\twentysevn$$

For the rotating black string solution with only a nonzero axionic
charge \HORNE, the gyromagnetic ratio $g$ is $2 - v^2$. So in the
ultrarelativistic limit, when $v \rightarrow 1$, $g$
becomes equal to $1$. In the limit when $v \rightarrow 0$, the $g$
factor becomes equal to $2$.

The extremal limit of the solution corresponds to $m^2 = a^2$,
where both the horizons coincide, $\it{i.e.}\, r_+ = r_-$. The
angular velocity $\Omega$ at the horizon is given by,

$$\eqalign{\Omega & = - {g'_{t \phi}\over{g'_{\phi \phi}}}\bigm\vert_{r = r_+}
\cr
& = {a\over{2m [m + \sqrt{m^2 - a^2}]}({1 + A\over 2})}
[1 + {\beta^2\over{1 + A}}]\cr}\eqn\twentyeit$$

In the limit, $\alpha_1, \alpha_2 \rightarrow 0$, this reduces to
that of the expression for the Kerr solution.
In the limit $\alpha_2 \rightarrow 0$, it reduces to that of the
four dimensional rotating black hole solution of Sen. In the limit
when $a \rightarrow 0$, the angular
velocity goes to zero. In the extremal limit ($m^2 = a^2$), $\Omega$
goes as ${1\over{2 a}}$ along with quantities depending on $\alpha_1$,
$\alpha_2$. Now if one considers the $a \rightarrow 0$ limit, we find that
$\Omega$ diverges. This is also true for the electrically neutral
rotating black string solution.

We also compute the surface gravity of the black string which is
given by,

$$\eqalign{\kappa &= lim_{r \rightarrow r_+} {\sqrt{g'^{r r}}}
\partial_r {\sqrt{- g'_{t t}}}\bigm\vert_{\theta = 0} \cr
 &= {\sqrt{(m^2 - a^2)}\over{2 m [m + \sqrt{m^2 - a^2}]}
 ({1 + A\over 2})} [1 + {\beta^2\over{1 + A}}]\cr}\eqn\twentynine$$

 This expression shows that in the extremal limit,
 surface gravity goes to zero. In the limit $a \rightarrow 0$,
 surface gravity is proportional to ${1\over 4 m}$. Infact it is
 given by,

 $$\kappa = {1\over{4 m}({1 + A\over 2})} [1 + {\beta^2\over{1 + A}}]
 \eqn\thirty$$

 In the limit when $\alpha_1$, $\alpha_2$ $\rightarrow 0$, it
 reduces to that of the expression for the Kerr solution and when
 $\alpha_2$ $\rightarrow 0$, it reduces to that of the four dimensional
 rotating black hole solution.
 The Hawking temperature can be calculated using the
 relation, $T = {\kappa\over{2 \pi}}$. For the rotating black string,
 Hawking temperature goes to zero in the extremal limit unlike the
 nonrotating case, where it diverges in the extremal limit.

 To summarize, in this paper we have constructed the most general
 electrically charged rotating black string solution in the five
 dimensional low  energy heterotic string theory using the powerful method
 of twisting, which allows us to generate new nontrivial solutions
 from the known ones. We have also studied the extremal limit of
 this solution carrying mass, electric charge, axion charge
 and angular momentum per unit length. The angular momentum was found to
 dominate over the charge in the extremal limit. The extremal limit
 is basically independent of $Q$. The extremal
 nonrotating black string corresponds to the fundamental string itself
 and is boost invariant in the $x - t$ plane \HORO. This was also
 shown to be true
 for the nonrotating charged black string \MICRO. Rotating black string
 is quite different and we do not know whether in the extremal
 limit it can be viewed as the
 field outside the fundamental string as we expect the solution for
 the fundamental string to be spherically symmetric.
 The study of black strings and p-branes is certainly interesting
 because of their close relationship with the string soliton solution
 as well as the fundamental strings themselves. Also, it has been
 shown in ref. \HHS\ that the dual solutions describing the extremal
 black strings are equivalent to plane fronted waves, which means that
 the corresponding metric describes a string moving at the speed of
 light. One would like to ask similar questions in the case where
 both rotation and charge are present. The other question we would
 like to ask is regarding the stability of these solutions.
 A more detailed study of black strings and p-branes
 will hopefully shed some light on these issues.

 \noindent{\bf {Acknowledgement}:} I would like to thank Ashoke Sen and
  S. Fawad Hassan for many useful discussions.

\refout

\end